\documentclass[preprint,showpacs,aps,preprintnumbers,amsmath,amssymb,tightenlines]{revtex4}

\usepackage{graphicx}
\usepackage{dcolumn}
\usepackage{bm}

\providecommand{\beqa}{\begin{eqnarray}}
\providecommand{\eeqa}{\end{eqnarray}}
\providecommand{\phid}{{\dot \phi}}
\providecommand{\Xd}{{\dot X}}

\providecommand{\cO}{{\cal O}}
\providecommand{\mt}{{\tilde m}}

\begin{document}

\preprint{UMD-PP-05-019} \preprint{BA-04-10}

\title{Ghost Cosmology: \\Exact Solutions, Transitions Between Standard
Cosmologies and Ghost Dark Energy/Matter Evolution}

\author{Axel Krause}
\email{krause@physics.umd.edu}
\affiliation{Department of Physics, University of Maryland,\\
College Park, Maryland 20742, USA}

\author{Siew-Phang Ng}
\email{spng@bartol.udel.edu}
\affiliation{Department of Physics, University of Maryland,\\
College Park, Maryland 20742, USA\\ and \\
Bartol Research Institute, University of Delaware,\\
Newark, Delaware 19716, USA}

\date{\today}

\begin{abstract} 
The recently proposed infrared modification of gravity
through the introduction of a ghost scalar field results in a number of
interesting cosmological and phenomenological implications. In this paper,
we derive the exact cosmological solutions for a number of scenarios where
at late stages, the ghost behaves like dark matter, or dark energy. The
full solutions give valuable information about the non-linear regime
beyond the asymptotic first order analysis presented in the literature.
The generic feature is that these ghost cosmologies give rise to smooth
transitions between radiation dominated phases (or more general power-law
expansions) at early epochs and ghost dark matter resp.~ghost dark energy
dominated late epochs. The current age of our universe places us right at
the non-linear transition phase. By studying the evolution backwards in
time, we find that the dominance of the ghost over ordinary baryonic
matter and radiative contributions persists back to the earliest times
such that the Friedmann-Robertson-Walker geometry is dictated to a good
approximation by the ghost alone. We also find that the Jeans instability
occurs in the ghost dark energy scenario at late times, while it is
absent in the ghost dark matter scenario. 
\end{abstract}


\maketitle

\section{Introduction}

An interesting proposal to explain dark matter or dark energy by
an infrared modification of standard gravity was made recently in
\cite{ACLM} (there is a rather long history of earlier attempts at
infrared modifications of gravity, see e.g.~\cite{His}). The key
idea is that gravity at the present epoch might be in a Higgs
phase. This is achieved by the introduction of an additional ghost
scalar field, $\phi$, with the wrong-sign kinetic term. If the
scalar has a vanishing vacuum expectation value (vev), one would
obviously obtain an unstable vacuum since the kinetic energy would
be unbounded from below. However, a stable vacuum can emerge if
one allows for a more general kinetic term. The ghost scalar might
then condense in a vacuum with non-trivial constant velocity in
field space
\beqa \phi(t) = \pm\sqrt{C}t \; .
\label{vev}
\eeqa
Notice that this vacuum breaks Lorentz invariance spontaneously by
breaking time translation invariance. The ghost's kinetic term is
described by a kinetic function $P(X)$ where \footnote{ Opposite
to \cite{ACLM} we are using in this paper a metric signature
$(-,+,\hdots,+)$ and the general relativity conventions of
\cite{HE}.} \beqa X = -\partial_\mu \phi
\partial^\mu \phi
\label{StanKinFun} \eeqa is (up to a factor two) the standard kinetic
term. The interesting observation is then that, once the ghost has
condensed, it assumes the equation of state of either matter or
vacuum energy. Consequently it represents an interesting candidate
for both dark matter and dark energy \cite{ACLM}.

To study fully the cosmological consequences of the ghost
condensation proposal one first needs to derive the corresponding
cosmological solutions. Though asymptotic analysis at late stages,
close to the condensation point, does already provide important
insights, it cannot address the full history of the ghost's
evolution. In particular one might ask what happens to the ghost's
mass density when we adjust it, according to the proposal of
\cite{ACLM}, at the current epoch with the observed dark
matter/energy density. Will the ghost still dominate at earlier
times or become negligible as compared to ordinary matter
resp.~radiation? How is the scale-factor of the universe affected
by the ghost at earlier times? Moreover, with the current age of
the universe, as given by its Hubble time
$H_0^{-1}=(10^{-33}\text{eV})^{-1}$, has the ghost already had
sufficient time to condense, or are we still far away from this
critical point. Moreover, one would like to know when the
asymptotic behavior sets in and can be trusted as an approximate
solution.

Our goal in this paper is to provide these ghost cosmological
solutions for a number of cases, as a starting point for more
detailed cosmological and astrophysical investigations of the
ghost condensation proposal. We will see that a ghost cosmology
leads generically to a universe which evolves from a fractional
power-law expansion (for the simplest model with quadratic $P(X)$
one finds a $t^{1/2}$ radiative phase; for models based on cubic,
quartic or higher power $P(X)$ more exotic fractional power time
dependences arise) to a late time matter or dark energy dominated
phase. A ghost cosmology defined through a particular $P(X)$ will
therefore smoothly interpolate between two types of standard
cosmologies. This in itself seems to be of technical interest,
given that normally when going from one epoch to the next, one
simply glues standard cosmologies together in a continuous but
non-smooth fashion. Furthermore, we will see that at the current
epoch, if the ghost indeed accounts for a sizeable contribution to
dark matter/energy, neither of the asymptotic late or early time
limiting cosmological solutions applies. Instead, we find
ourselves right in the highly non-linear transitory stage between
early and late time asymptotics. Hence the ghost condensation
process, which is shown to endure for an infinite amount of time,
is far from being finished. Via the ghost cosmological solutions
the ghost's mass density will depend in a non-trivial way on
cosmic time. In particular, it will turn out that if the current
ghost's mass density is of the size of the presently observed dark
matter/energy density, then the ghost will always have dominated
other contributions like ordinary matter or radiation in the past
and will continue to do so in the future. Thus as a good
approximation one can neglect ordinary matter and radiation at
other epochs as well and get a sufficiently accurate cosmological
description by just considering the ghost-gravity system alone.
Our thus derived ghost cosmologies are therefore relevant
not only at the current epoch but also at earlier and later
epochs. One interesting aspect of these ghost cosmological
solutions concerns the coincidence problem. In general, since the
ghost's contribution to the dark energy density does not stay
constant in time one would expect not only the dark matter density
but now also the ghost's dark energy density to increase towards
earlier times. This should at least lead to some alleviation of
the coincidence problem and would be interesting to study in
future work.

Finally, it would be nice to embed the ghost condensation proposal
into M-theory. The ghost, which appears only derivatively coupled,
satisfies a shift symmetry and could therefore easily be
identified as an axion. The hope would be that M-theory can fix
the magnitude of the ghost dark matter or energy which in the
effective field theory framework needs to be adjusted to the
observed values. Moreover M-theory predicts dark matter/dark
energy on its own (see e.g.~\cite{MDM}) and it would be
interesting to see the interplay of these contributions with the
ghost contribution. For other interesting implications of the
ghost condensate proposal or related subjects, see the references in
\cite{Others}.

Without further a due, we will start the next section by
presenting the general framework for cosmology with the ghost
scalar. Section III will be devoted to the full cosmological
solution related to the appearance of ghost dark matter at late
epochs. Here, as we have alluded to already, we find a transition
from a radiation dominated early epoch to a ghost dark matter
dominated late epoch. Section IV gives the cosmological solution
for $P(X)=X^n$ which arises in any Taylor-expansion of a more
complicated kinetic function and is in particular relevant at
early times. The outcome is a simple fractional power-law time
dependence. Section V derives the cosmological solutions which
give rise to a late time (ghost) dark energy accelerated
expansion. The general result is that the full solution
interpolates between an early epoch radiation dominated phase and
the mentioned (ghost) dark energy dominated late epoch. In Section
VI, we discuss the possibility of a transition between an early
period with large vacuum energy, suitable for instance to sustain
inflation, and a late period with a smaller amount of dark energy.
Section VII includes ordinary baryonic matter and radiation in
addition to the ghost. We study their evolution and find that if
the ghost accounts for all of dark matter resp.~dark energy
observed today that the dominance of the ghost will exist not only
today but persists back to much earlier times. For the evolution
of the scale-factor of the Friedmann-Robertson-Walker (FRW)
geometry one can therefore neglect other sources besides the ghost
to a very good approximation. The final section VIII analyzes the
potential Jeans instability of our cosmological solutions.

\section{Ghost Cosmology}

In this section, we will present the general framework describing
the cosmological evolution of a FRW spacetime in the framework of
general relativity coupled to a ghost scalar. In later sections,
we will then apply this formalism to concrete cases covering late
time generation of ghost dark matter or dark energy.

\subsection{The Cosmological Equations}

Following the proposal of \cite{ACLM}, one replaces the usual
kinetic term $X$ for the ghost scalar $\phi$ through some more
complicated function $P(X)$ leading to a ghost gravity theory
defined by an action \beqa S = \int d^4x \sqrt{-g} \left(
\frac{R}{16\pi G} + M^4 P(X) \right) \; . \label{action1} \eeqa
Notice that it is not necessary to add a cosmological constant
term separately by hand as it can be absorbed into a constant
added to $P(X)$. We will elaborate more on this issue later in
section V. The mass scale $M$ is not determined {\em a priori} by
the theory itself but should be regarded as a free parameter which
needs to be fixed by phenomenological considerations. The range of
$M$ becomes thus confined to \cite{ACLM} \beqa 1 \text{meV}\le M
\le 10 \text{MeV} \eeqa where the upper bound can possibly even be
relaxed to the GeV regime \cite{Markus}. By variation of this
action w.r.t.~the metric and $\phi$, we obtain the Einstein
equation \beqa G_{\mu\nu} = 8\pi G T_{\phi,\mu\nu}
\label{Einstein} \eeqa with ghost energy-momentum tensor \beqa
T_{\phi,\mu\nu} &= M^4(P(X)g_{\mu\nu} +
2P'(X)\partial_\mu\phi\partial_\nu\phi) \label{EMT1} \eeqa and the
equation of motion for $\phi$ \beqa
\partial_\mu(\sqrt{-g}P'(X)\partial^\mu\phi) = 0 \; .
\eeqa For the ghost condensate proposal, it is essential that
$P(X)$ exhibits a minimum at some positive value $C$ such that $X$
becomes timelike at this value and consequently $\phi$ acquires a
non-vanishing time-dependent vev (\ref{vev}) in this vacuum state.
A stability analysis of fluctuations around the condensate point
implies the constraints \cite{ACLM} \beqa P'(C) \ge 0 \; , \qquad
P'(C)+2CP''(C) \ge 0 \; . \label{Constraint} \eeqa Besides these
requirements, $P(X)$ is {\em a priori} arbitrary. Though it will
soon turn out that the Friedmann equation (for the case with flat
spatial sections studied in this paper) imposes a further
consistency constraint, it is typically the case that this
additional constraint is, for the cases of $P(X)$ studied in this
paper, satisfied wherever (\ref{Constraint}) holds true.

The ghost condensate breaks time translation invariance and the
shift symmetry in $\phi$ down to a diagonal subgroup \cite{ACLM}
while leaving all other spatial symmetries like e.g.~rotation
invariance intact. A spatially homogeneous and isotropic universe
is therefore compatible with the idea of a ghost condensate and
will be adopted throughout this paper. The corresponding geometry
is described by an FRW metric where our focus is on the case with
flat 3d sections \beqa ds^2 = -dt^2 + a^2(t) \left(
dr^2+r^2(d\theta^2+\sin^2\theta d\phi^2) \right) \; . \label{FRW}
\eeqa In compliance with the homogeneity and isotropy assumption,
the ghost scalar cannot depend on the spatial coordinates but only
on time \beqa \phi=\phi(t) \; . \eeqa We therefore see that in the
FRW background $X$ depends only on the ghost's time derivative
\beqa X = \phid^2 \; . \label{X} \eeqa Consequently the equation
of motion for $\phi$ simplifies to \beqa
\frac{d}{dt}\big(a^3(t)\phid P'(X)\big) = 0 \; , \label{GhostEOM}
\eeqa which is solved by \beqa a^3(t)\phid P'(X) = \text{const} \;
. \label{GhostEOM2} \eeqa If at late times we want to have an ever
increasing scale-factor $a\rightarrow\infty$ then either $\phid$
or $P'(X)$ have to vanish. The former choice, which would amount
to redshifting the ghost velocity to zero, does not take us to the
ghost condensate vacuum state and will not be of interest to us
here. We will therefore adopt the second boundary condition,
namely that asymptotically, \beqa X \rightarrow C \quad
\text{with} \quad P'(C)=0 \qquad\; \text{as} \qquad\; t
\rightarrow \infty \; , \label{BC} \eeqa which describes the
condensation of the ghost $\phi(t) \rightarrow \pm \sqrt{C}t$.

This boundary condition implies that the condensation process
takes place in an infinite amount of time. In view of the fact
that the condensation point $X=C$ for the simplest quadratic
kinetic function $P(X)=\frac{1}{2}(X-C)^2$ marks the border
between stable and unstable states \cite{ACLM} this is what we
want. Otherwise, one would have to be concerned about
`overshooting' into the unstable regime when $\phi$ condenses.
Nevertheless, one might also wonder whether it might be possible
for the condensation process to finish in a finite amount of time.
If we assume this to be true then we learn from
Eq.~(\ref{GhostEOM2}) that $a(t_c)=\infty$ would have to hold at
finite time $t_c$ indicating that the solution possesses a
singularity which can be reached in finite time and is
geodesically incomplete. We will discard such solutions and are
hence forced to assume that the condensation process terminates at
$t_c=\infty$.

From the Einstein equation, two independent equations are
obtained. Thus, it might seem that the problem might be
overdetermined: three equations (two from the Einstein equation
plus the ghost equation of motion) for just two unknown functions
(the scale-factor $a(t)$ and the ghost $\phi(t)$). This is,
however, not true. To this end, let us note that \beqa
\triangledown^\mu T^\phi_{\mu\nu} = -\delta_\nu^0
\frac{d}{dt}(a^3\sqrt{X}P') \; , \eeqa where we have used that
$\dot{P} = P'\Xd$. Hence, local energy conservation is equivalent
to the equation of motion for the ghost given earlier in
Eq.~(\ref{GhostEOM}). \beqa \triangledown^\mu T^\phi_{\mu\nu} = 0
\qquad \Leftrightarrow \qquad \text{EOM for }\phi \; . \label{Equ}
\eeqa Since local energy conservation is already implied by the
Einstein equation, the same is true for the ghost equation of
motion. We are therefore left with just two independent equations
coming from the Einstein equation for the two unknown functions
$a(t),X(t)$. By noticing that the ghost depends only on the time
coordinate, we can always replace its time derivative by
$\sqrt{X}$ and thus work exclusively with the more convenient
variable $X$. This would no longer be possible if we allow for an
inhomogeneous or anisotropic universe as in these cases, the ghost
would depend on more than just one coordinate.

Having dealt with the scalar equation of motion, let us now
consider the equations for the metric. For the FRW background, the
Einstein equation leads to two independent equations, the {\em
Friedmann equation} \beqa H^2 = \frac{m^2}{3} (2 X P'(X) - P(X))
\label{Fried} \eeqa and the evolution equation \beqa
H^2+2\frac{\ddot{a}}{a} = - m^2 P(X) \eeqa where, as usual, the
Hubble parameter $H$ denotes ${\dot a}/a$ and we have defined the
mass-parameter $m$ through \beqa m = \sqrt{8\pi G} M^2 =
\frac{M^2}{M_{Pl}} \; , \eeqa with $M_{Pl}$ the reduced Planck
mass. Like the ``see-saw'' relation of Grand Unified Theories
employed to explain the smallness of the neutrino masses
\cite{Seesaw}, we have here a similar ``see-saw'' relation where
$M$ is the geometric mean of $m$ and $M_{Pl}$. With a range of $M$
between $1$meV and $1$GeV the parameter $m$ lies between the
extremely small energies \beqa 4\times 10^{-34}\text{eV} < m <
4\times 10^{-10}\text{eV} \; . \eeqa Notice that the lower bound
corresponds to the current Hubble scale which is $H_0\simeq
10^{-33}\,$eV.

Combining the evolution and the Friedmann equation to eliminate
$H$ from the former yields the {\em acceleration equation}
\beqa \frac{\ddot{a}}{a} = -\frac{m^2}{3}(X P'(X) + P(X))
\label{Accel}
\eeqa which is more convenient to use than the evolution equation.

\subsection{The General Formal Solution to the Cosmological Equations}

The Friedmann equation (for the 3d flat case with $k=0$ considered
here), due to the positivity of $H^2$, imposes another consistency
constraint \beqa 2 X P'(X) \ge P(X) \; , \quad \forall \; X(t)
\label{PosCon} \eeqa on any candidate ghost kinetic function
$P(X)$. We will see shortly that this constraint is equivalent to
demanding a non-negative gravitating mass density $\rho_\phi$
coming from the ghost. It should, however, be noted that further
sources of positive mass densities or a negative 3d curvature
would contribute positively to $H^2$, thereby relaxing the above
constraint somewhat. Let us now proceed by formally solving the
Friedmann and the acceleration equation.

First, one observes that the Friedmann equation, a first order
differential equation for $a(t)$, can be formally solved in terms
of $X$ through ($a_i = a(t_i)$) \beqa a(t) = a_i e^{f(t,t_i)} \; ,
\quad f(t,t_i) = \pm \frac{m}{\sqrt{3}}\int_{t_i}^t dt'\sqrt{2
X(t') P'(X(t'))-P(X(t'))} \label{Scale} \eeqa with $t_i$ an
initial reference time. Now, we can plug this solution into the
lhs of the acceleration equation to eliminate the scale-factor and
obtain the following first order differential equation for $X(t)$
\beqa 2 \dot{(P'(X))} X + \dot{P} = \mp 2\sqrt{3}m X P'(X)
\sqrt{2XP'(X)-P(X)} \; , \eeqa which can also be written more
compactly as \beqa \frac{d}{dt}\ln\left|(P'(X))^2 X \right| = \mp
2\sqrt{3}m \sqrt{2XP'(X)-P(X)} \; . \label{XODE} \eeqa This is
then the equation, which when supplemented with the boundary
condition (\ref{BC}), determines $X(t)$. Assuming that $X(t)$ has
been found by solving (\ref{XODE}) we can then evaluate explicitly
the function $f(t,t_i)$ and thus determine the scale-factor from
(\ref{Scale}). Indeed, with help of (\ref{XODE}) we can show for
general $P(X)$ that \beqa f(t,t_i) = -\frac{1}{6}\int_{t_i}^t dt'
\frac{d}{dt'} \ln\left|X (P'(X))^2\right| = \ln
\left|\left(\frac{X_i (P'(X_i))^2}{X(t) (P'(X(t))^2}\right)^{1/6}
\right| \; , \eeqa which gives us for the scale-factor \beqa a(t)
= a_i \left|\frac{X_i (P'(X_i))^2}{X(t) (P'(X(t))^2}\right|^{1/6}
\; . \label{aSol} \eeqa As can be easily verified, this solution
for $a(t)$ satisfies (\ref{GhostEOM2}) and therefore is a solution
to the ghost equation of motion. This is in accordance with what
we said earlier about the redundancy of the ghost equation of
motion as it is already implied by the Einstein equation.

In the following, we will solve (\ref{XODE}) for three classes of
functions $P(X)$ and study the corresponding ghost cosmologies.
But before doing so, let us briefly return to the energy-momentum
tensor (\ref{EMT1}). In general the diagonal components of the
energy-momentum tensor represent the mass density and pressure,
$T_{\mu\nu} = (\rho , p g_{rr}, p g_{\theta\theta}, p
g_{\phi\phi})$. By comparison to (\ref{EMT1}) one may therefore
also infer a pressure and a mass density associated with the ghost
field \beqa \rho_\phi = M^4(2 X P'(X) - P(X)) \; , \qquad p_\phi =
M^4 P(X) \; . \label{rhop} \eeqa These quantities will turn out to
be very useful in understanding various limits of our cosmological
solutions. The constraint (\ref{PosCon}) implies that $\rho_\phi$
can never become negative.

\section{Ghost Dark Matter Case: Transition from Radiation- to
Matter-Dominated Universe with $P(X)=\frac{1}{2}(X-C)^2$}

Let us now apply the formalism of the previous section to some
concrete models. The simplest choice for $P(X)$ letting the ghost
condense at some finite positive value $X = C$ is given by the
parabola \beqa P(X) = \frac{1}{2}(X-C)^2 \; . \label{P1} \eeqa The
model defined by this kinetic function will be relevant for the
question of ghost dark matter at the present epoch as we will see.
Both the stability criterion (\ref{Constraint}) and the constraint
(\ref{PosCon}) require that we work in the regime where $X\ge
C>0$. Notice that in the excluded regime of negative $X$, it would
be required by (\ref{X}) that we work with a complex ghost field
while for positive $X$, it is consistent to work with a real
field.

The Friedmann and acceleration equation become
\begin{alignat}{3}
H^2 &= \frac{m^2}{6}(X-C)(3X+C)
\label{HDM} \\
\frac{\ddot{a}}{a} &= -\frac{m^2}{6}(X-C)(3X-C) \; ,
\end{alignat}
and it follows from the acceleration equation that any
cosmological solution must describe a {\em decelerated expansion}
for which \beqa \ddot{a}\le 0 \; . \eeqa where equality is
attained when the ghost reaches the condensation point
$X\rightarrow C$ at late times.

The differential equation (\ref{XODE}) determining $X(t)$ can be
written most conveniently as \beqa \Xd =
\mp\sqrt{6}m\frac{X(X-C)^{3/2}(3X+C)^{1/2}}{3X-C} \eeqa and gets
solved by the following implicit expression for $X(t)$ \beqa
Q(Y)+2\arctan Q(Y)-Q_i-2\arctan Q_i = \pm\sqrt{6}Cm(t-t_i)
\label{Y1} \eeqa where \beqa Q(Y) = \sqrt{\frac{3Y(t)+1}{Y(t)-1}}
\; , \qquad Y(t)=\frac{X(t)}{C} \; . \eeqa and $Q_i=Q(t_i)$
denotes the value at initial time $t_i$. Subsequently, we will
choose the solution with positive sign leading to a time flow in
positive direction while the ghost scalar condenses. To determine
its value we impose the following initial boundary condition \beqa
X(t_i) \gg C, \;\; Y(t_i) \gg 1  \qquad \text{as} \qquad t
\rightarrow t_i = 0 \; , \eeqa which fixes the integration
constant with \beqa Q_i = \sqrt{3} \; . \eeqa Notice that at late
times where the boundary condition must be (\ref{BC}), it cannot
be determined as it drops out in this limit. This solution for
$X(t)$ maps the time interval $t_i=0 \le t \le \infty$ faithfully
into the $X$ interval $\infty \ge X(t) \ge C$. {\em The ghost
condensation process is therefore seen to take an infinite amount
of time}. Therefore, once the cosmological evolution starts within
the regime where $X>C$, we don't have to worry that we might be
driven towards the unstable regime where $X<C$ in finite time. The
time-dependence of the FRW scale-factor follows from (\ref{aSol})
as \beqa a(t) \propto \frac{1}{X(t)^{1/6}(X(t)-C)^{1/3}} \propto
\frac{1}{Y(t)^{1/6}(Y(t)-1)^{1/3}} \; . \label{a1} \eeqa

Before we discuss this solution, let us first look into its early
and late time behavior. Close to the initial time where \beqa t
\rightarrow 0 \; , \quad Y \rightarrow \infty \; , \eeqa we can
expand the solution for $Y(t)$ and $a(t)$ and find
\begin{alignat}{3}
Y(t) &= \frac{1}{\sqrt{2}C m t} \\
a(t) &\propto (C m t)^{1/2} \; .
\end{alignat}
Hence at early times, the ghost scalar behaves like radiation and
we obtain the characteristic $a(t)\propto t^{1/2}$ behavior of a
{\em radiation-dominated type universe}.

On the other hand at late times when the ghost has nearly reached
its condensation point ($Y$ approaches $1$ from above)
\beqa
t \rightarrow \infty \; , \quad Y \rightarrow 1 \; ,
\eeqa
expansion of the solution leads to the asymptotic solution
\begin{alignat}{3}
Y(t) &= 1 + \frac{2}{3(C m t)^2} \\
a(t) &\propto (C m t)^{2/3} \; .
\end{alignat}
Thus, as expected, the ghost field behaves at late times like
non-relativistic matter with the characteristic $a(t)\propto
t^{2/3}$ behavior. It could therefore account for some or all of
dark matter at the present epoch. The time evolution of this type
of ghost dark matter will be investigated in section VII. The {\em
full cosmological solution (\ref{Y1}),(\ref{a1}) therefore
describes a smooth transition from a radiation dominated early
epoch to a ghost dark matter dominated late epoch}.

\begin{figure}[th]
\includegraphics[scale=0.7]{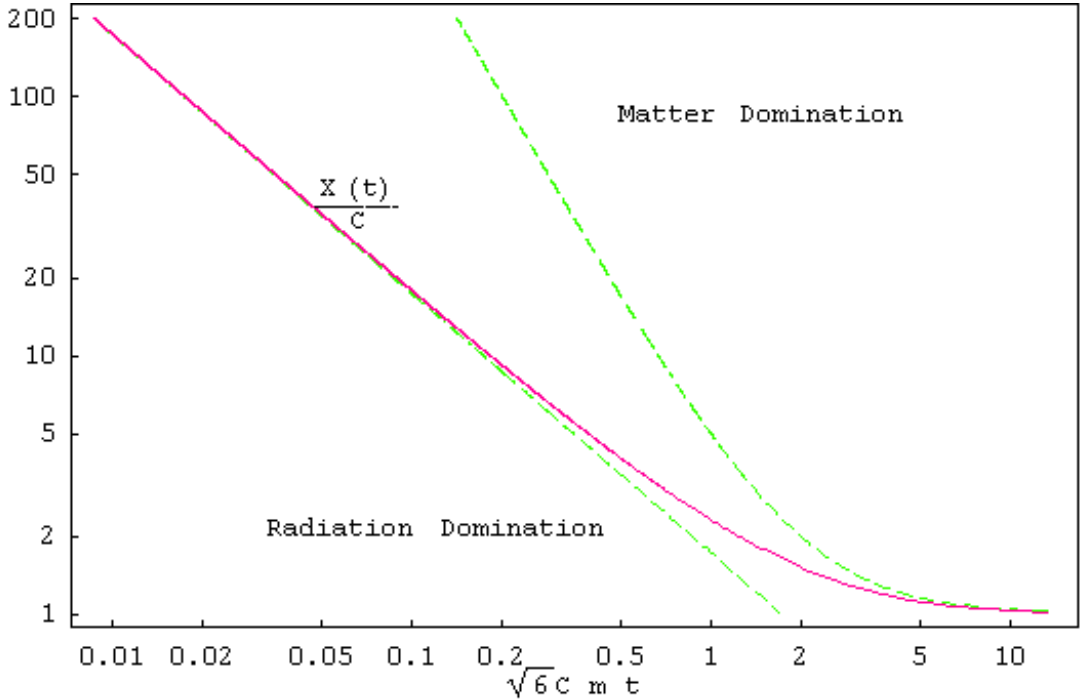}
\caption{\label{FigYPlot1} Plot of the ghost field $X(t)$ (red
middle curve) as a function of time for a theory based on
$P(X)=\frac{1}{2}(X-C)^2$. It is clearly visible how $X(t)$
interpolates between the early time radiation dominated behavior
(green lower dashed curve) and the late time ghost dark matter
dominated behavior (green upper dashed curve). The transition
occurs during the relatively short time interval $\sqrt{6}C m
t\sim 0.1-10$.}
\end{figure}

\begin{figure}[th]
\includegraphics[scale=0.7]{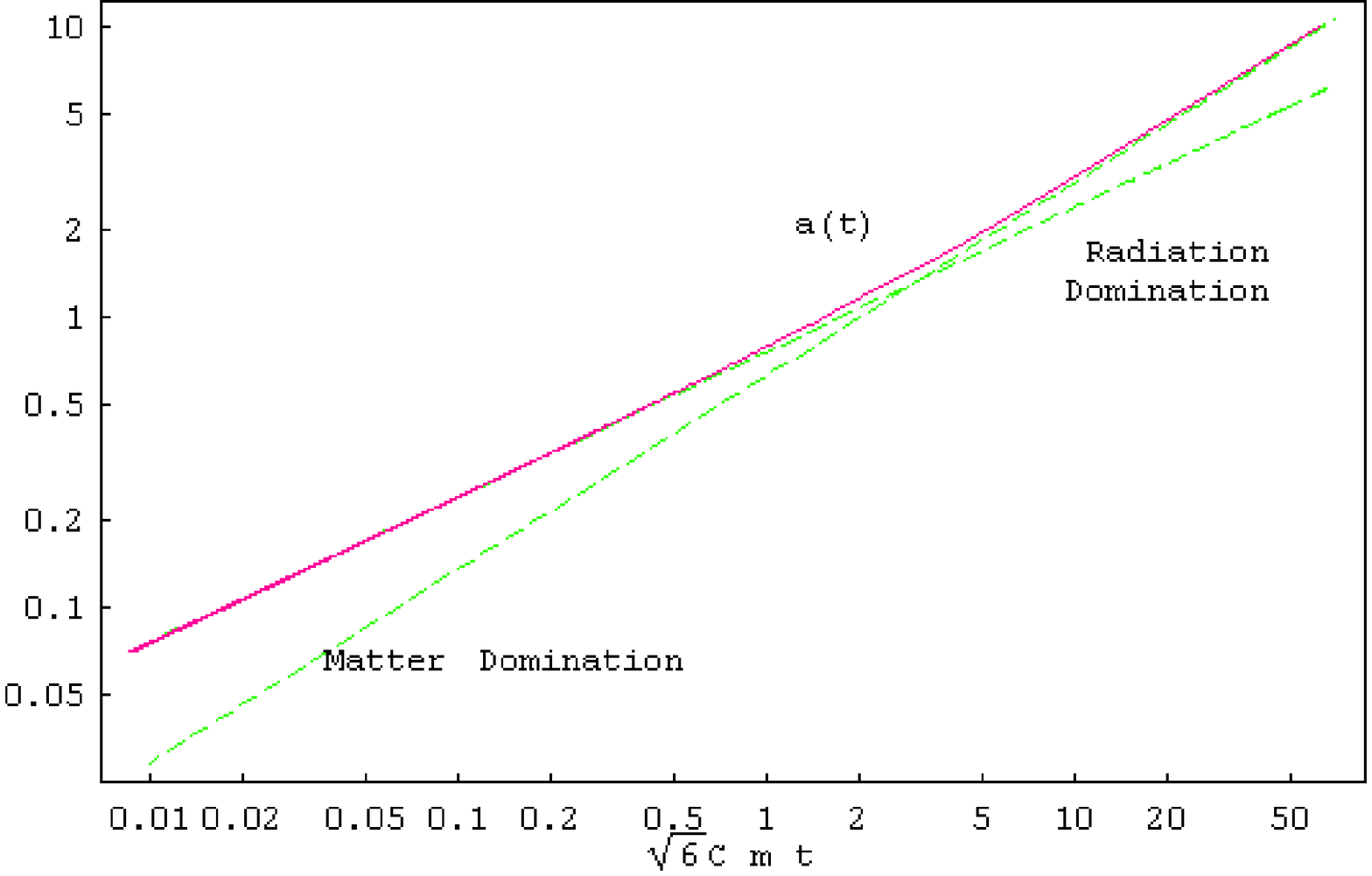}
\caption{\label{FigaPlot1} Plot of the scale factor $a(t)$ as a
function of time for a ghost cosmology with
$P(X)=\frac{1}{2}(X-C)^2$. The scale factor (red middle curve)
evolves from a $t^{1/2}$ radiation dominated behavior at early
times (green upper dashed curve) to a $t^{2/3}$ ghost dark matter
dominated behavior at late times (green lower dashed curve).}
\end{figure}
This transition from a radiation- to a matter-dominated universe
can also be understood by inspection of the ghost's matter density
$\rho_\phi$ and pressure $p_\phi$ at early and late times. For
$t\rightarrow 0, X(t)\gg C$ the expressions (\ref{rhop}), with
$P(X)$ given in (\ref{P1}), yield \beqa \rho_\phi \rightarrow 3
p_\phi \eeqa while at late times towards the end of the
condensation process, $t\rightarrow t_c=\infty, X\rightarrow C$ we
obtain \beqa \rho_\phi = 2M^4 C (X-C) + \cO((X-C)^2) \gg p_\phi =
\cO((X-C)^2) \; . \label{eos1} \eeqa Hence, the ghost field
assumes indeed the correct equation of state to behave as
radiation at an early epoch and non-relativistic matter at a late
epoch.

We will see later in section VII that with the current age of the
universe $t_0 = 1/H_0$ we are led to $\sqrt{6}C m t_0 \sim 1$. If
we now plot the full solutions for $X(t)$ (Fig.\ref{FigYPlot1})
and $a(t)$ (Fig.\ref{FigaPlot1}) they show besides the discussed
transition from a radiation to a matter dominated cosmology that
we are {\em currently right in the transition period}. As an
explanation of dark matter through the ghost, this is promising as
we see from (\ref{eos1}) that towards the very end of the
condensation process $\rho_\phi$ will vanish and hence its
contribution to non-relativistic matter as well.

\section{Behavior at Large $X$ and the General Case:
$P(X)=X^n$ and $P(X)=(X-C)^n$}

Exact cosmological solutions for more complicated $P(X)$ can
become quite involved. It is therefore important to understand the
general characteristics. Let us now concentrate on the
cosmological evolution when $X\gg C$, i.e.~at early times when the
ghost is still far away from its condensation point. Obviously in
this regime, any sufficiently smooth function $P(X)$ can be
Taylor-expanded in terms of powers of $X$. Let us therefore study
first the dynamics of just the simple power-law \beqa P(X)=X^n \;
, \label{P2} \eeqa which would occur in the Taylor expansion of
more complicated function $P(X)$. In this case the differential
equation for $X(t)$ is easily solved through \beqa
X(t)=\left(\pm\frac{\sqrt{3}m n}{\sqrt{2n-1}}t +
X_i^{-\frac{n}{2}}\right)^{-2/n} \; , \eeqa with initial value
$X_i=X(t_i)$ at initial time $t_i=0$. If we assume that initially
\beqa X_i \gg 1  \; , \eeqa and take the solution with positive
sign, then the result is a simple power dependence for $X(t)$ as
well as for $a(t)$ (by plugging $X(t)$ into the general
(\ref{aSol}))
\begin{alignat}{3}
X(t) &=\left(\frac{\sqrt{2n-1}}{\sqrt{3}m n}\right)^{2/n}
\frac{1}{t^{2/n}} \\
a(t) &\propto t^{\frac{2n-1}{3n}} \; .
\label{a2}
\end{alignat}

Indeed, it is easy to understand how this time-dependence arises.
Namely, for $P(X)=X^n$ we find that the mass density and pressure
of the ghost obey \beqa \rho_\phi = (2n-1) p_\phi \; . \eeqa
Expressed in terms of the usual equation of state $p=w\rho$, this
means that $w_\phi=\frac{1}{2n-1}$. Given that for a FRW universe
with flat spatial sections and constant equation of state
parameter $w$ one gets \beqa a(t) \propto t^{\frac{2}{3(1+w)}} \;
, \eeqa we see that with $w_\phi$ the solution (\ref{a2})
immediately follows. As a particular case, we recover the
radiation dominated universe if $n=2$.

The more complicated kinetic function \beqa P(X) = (X-C)^n
\label{P3} \eeqa which would, in contrast to the approximate
$P(X)=X^n$, correctly describe a condensation process and
generalize the simple parabolic $P(X)$ of the last section gives
rise to the following equation for $Y(t)$ \beqa
\big[(2n-1)Y-1\big]\dot{Y} = \mp (2\sqrt{3}C^{\frac{n}{2}}m) Y
(Y-1)^{\frac{n+1}{2}}\big[(2n-1)Y+1\big]^{\frac{1}{2}} \eeqa which
has a solution in terms of a combination of two Appell
hypergeometric functions. Since this solution is rather involved,
we will not present it here but confine ourselves to a discussion
of the early and late time behavior. During the initial stages we
have approximately a $P(X)=X^n$ behavior with the power-law
expansion (\ref{a2}) while at late stages close to the ghost's
condensation point when $X(t) \rightarrow C$ we find a ghost
equation of state \beqa \frac{p_\phi}{\rho_\phi} = (X-C)/2nC
\rightarrow 0 \eeqa indicating a matter dominated period which
once more might be used to describe dark matter. With the
generalization (\ref{P3}) to arbitrary powers, we therefore
recognize that {\em transitions from more general power-law
cosmologies $a(t)\propto t^{(2n-1)/3n}$ at an early epoch to a
late epoch ghost dark matter dominated $a(t)\propto t^{2/3}$
cosmology} are possible as well.

\section{Dark Energy Case: Transition from Radiation- to
De Sitter Universe with $P(X)=\frac{1}{2}(X-C)^2-D$}

Let us next see how one can accommodate for a dark energy
dominated de Sitter expansion phase at late times within the ghost
gravity framework. Adding an explicit cosmological constant
$\Lambda$ to the ghost-gravity action leads to \footnote{Again we
follow the conventions of \cite{HE}.}
\beqa
S = \int d^4x \sqrt{-g}
\left( \frac{(R-2\Lambda)}{16\pi G} + M^4 P(X) \right) \; .
\label{action2}
\eeqa

>From this action and also from the ensuing Einstein equation \beqa
G_{\mu\nu} = 8\pi G T_{\phi,\mu\nu} - \Lambda g_{\mu\nu} \eeqa
with $T_{\phi,\mu\nu}$ as given in (\ref{EMT1}), it is obvious
that a cosmological constant term can be completely absorbed into
$P(X)$ simply by adding a constant \beqa P(C) =
-\frac{\Lambda}{m^2} \eeqa to the kinetic function,
$P(X)\rightarrow P(X)+P(C)$. We can therefore still work with the
simpler action (\ref{action1}) and use our formalism as presented
in section II. Unlike the simple parabolic $P(X)$ case which we
studied before in relation to dark matter which had $P(C)=0$, we
would now require a non-vanishing $P(C)\ne 0$ at the condensation
point $X=C$ which is characterized furthermore through $P'(C)=0$.
This will lead to either de Sitter (dS) or anti-de Sitter (AdS)
universes whenever such a critical point is reached
\begin{alignat}{3}
P(C) < 0 \quad &\Rightarrow \quad \Lambda > 0 \quad \Rightarrow
\quad \text{dS} \\
P(C) > 0 \quad &\Rightarrow \quad \Lambda < 0 \quad \Rightarrow
\quad \text{AdS}
\end{alignat}

The simplest theory exhibiting such a late time dS expansion
arises by adding to our previous parabolic $P(X)$ a positive
constant $D$
\beqa
P(X)=\frac{1}{2}(X-C)^2-D \; ,
\eeqa
such that at the critical point $P(X=C)=-D$. Upon condensation the
ghost then behaves like a positive cosmological constant
\beqa \Lambda = m^2 D \; . \eeqa
The Friedmann and acceleration equations for this specific case
become
\begin{alignat}{3}
H^2 &= \frac{m^2}{3}\Big(\frac{1}{2}(X-C)(3X+C)+D\Big)
\label{HDE} \\
\frac{\ddot{a}}{a} &= -\frac{m^2}{3}
\Big(\frac{1}{2}(X-C)(3X-C)-D\Big) \; .
\end{alignat}
It is obvious from the acceleration equation that {\em at early
times} where $X\gg C$ the ensuing ghost cosmology undergoes a
period of {\em deceleration}, $\ddot{a}<0$, while {\em at late
times} when the ghost condenses and we have $X\rightarrow C$ this
will turn into a period of {\em acceleration}, $\ddot{a}>0$.
Working once more in the regime where $X\ge C$ will ensure that
both constraints, (\ref{Constraint}) and (\ref{PosCon}), are
satisfied such that the vacua considered are stable and the ghost
mass density is always non-negative. This regime is also
compatible with the boundary condition (\ref{BC}) at late times.

To find the corresponding cosmological evolution we have to solve
the differential equation for $X(t)$ (\ref{XODE}). To this end, it
is most convenient to work with the rescaled variable
$Y(t)=X(t)/C$ and to introduce the parameters \beqa \mt =
\pm\sqrt{6} m C \; , \qquad d=\frac{2D}{C^2} \; . \eeqa The
differential equation (\ref{XODE}) then becomes \beqa \dot{Y}
(3Y-1) = -\mt Y(Y-1)\sqrt{(Y-1)(3Y+1)+d} \; . \eeqa Once a
solution to this equation has been found, the scale-factor follows
from (\ref{aSol}) as \beqa a(t) \propto \frac{1}{Y(t)^{1/6}
(Y(t)-1)^{1/3}} \; . \eeqa We will now present the solutions in
different classes according to the value $d$ which determines the
late period cosmological constant.

\subsection{$d>1, d\ne\frac{4}{3}$}

For the first case $d>1, d\ne\frac{4}{3}$ which comprises the case
with arbitrary large vacuum energy the differential equation
yields the solution
\beqa
\mt t = \pm \frac{1}{\sqrt{d-1}}
\ln\left|\frac{U_{\pm}(t)}{U_{\pm,i}}\right|
+ \frac{2}{\sqrt{d}}\ln\left|\frac{V(t)}{V_i}\right| \; ,
\eeqa
where
\begin{alignat}{3}
U_{\pm}(t) &= \frac{2}{Y(t)}\left(\frac{\pm(d-1-Y(t))}{\sqrt{d-1}}
+ \sqrt{3Y^2(t)-2Y(t)-1+d}\right) \; , \\
V(t) &= \frac{1}{(Y(t)-1)}\left(\frac{d+2(Y(t)-1)}{\sqrt{d}}
+ \sqrt{3Y^2(t)-2Y(t)-1+d}\right)
\end{alignat}
with initial values $U_{\pm,i},V_i$ at time $t_i=0$ and either
both plus signs or both minus signs have to be chosen. For $d=4/3$
the function $U_\pm(t)$ would vanish identically in the relevant
interval $1\le Y(t)<\infty$. We will therefore study this case
separately and concentrate here on the case with $d\ne 4/3$.
Moreover, we will focus on the solution with plus sign choice,
i.e.~$U_+$.

The first thing to notice is that both $U_\pm(t)$ and $V(t)$ have
no zero in the interval $1\le Y(t)<\infty$ and therefore do not
change sign. Moreover, both $|U_+(Y(t))|$ and $V(Y(t))$ decrease
monotonically with $Y$ over the interval $1\le Y < \infty$ and
cover the range \beqa \infty\ge V(Y)\ge
\frac{2}{\sqrt{d}}+\sqrt{3} \; , \qquad
2\left|\frac{d-2}{\sqrt{d-1}}+\sqrt{d}\right|\ge |U_+(Y)|\ge
2\left|\frac{1}{\sqrt{d-1}}-\sqrt{3}\right| \; . \eeqa We
therefore learn that $|\mt| t$ decreases monotonically with $Y$
over the interval $1\le Y < \infty$. Hence, time increases if the
{\em universe evolves from large to small $Y$ values}. So if we
let the universe start off at initial time $t_i=0$ at some value
$Y_i>1$ and condense at later time $t_c$ at value $Y(t_c)=1$, then
this evolution is described by \beqa |\mt| t =
\frac{1}{\sqrt{d-1}} \ln\left|\frac{U_+(t)}{U_{+,i}}\right| +
\frac{2}{\sqrt{d}} \ln\left|\frac{V(t)}{V_i}\right| \label{Sol1}
\eeqa By inverting this implicit solution one obtains the
scale-factor via (\ref{aSol}). Notice that at the condensation
point $t$ goes to infinity, thus \beqa t_c=\infty \eeqa and we see
again that {\em it takes an infinite amount of time for the ghost
to condense} which is why we do not have to worry about entering
the unstable regime where $X$ is smaller than $C$.

\subsection{$d=\frac{4}{3}$}

For the special case with $d=\frac{4}{3}$ which we have left out
in the previous analysis one finds the simple explicit solution
\beqa Y(t)=\left(1-e^{-\frac{\mt (t-t_i)}{\sqrt{3}}}
\Big(1-\frac{1}{Y_i}\Big)\right)^{-1} \label {Y4} \eeqa with value $Y_i>1$ at
initial time $t_i$. To have a solution in the stable regime $Y\ge
1$ we choose the positive sign for $\mt$. Again, we see that {\em
the condensation process takes an infinite amount of time} until
it reaches $Y(t_c=\infty)=1$. With $t_i=0$ the ensuing scale
factor becomes \beqa a(t) \propto e^{\frac{|\mt| t}{3\sqrt{3}}}
\sqrt{1-e^{-\frac{|\mt|t}{\sqrt{3}}} \Big(1-\frac{1}{Y_i}\Big)} \;
. \label{a4} \eeqa

\subsection{$d=1$}

For the case with $d=1$ the solution reads
\beqa
\mt t = 2\ln\left|\Big(\frac{W(t)+1}{W_i+1}\Big)
\Big(\frac{W_i-1}{W(t)-1}\Big)\right|
-(W(t)-W_i) \; ,
\eeqa
where
\beqa
W(t) = \sqrt{3}\left(\frac{\sqrt{1+3(3Y(t)^2-2Y(t))}-1}
{\sqrt{1+3(3Y(t)^2-2Y(t))}+1}\right)^{1/2}
\eeqa
and initially at $t_i=0$ we have $W(t_i)=W_i$.

Notice that the interval $1\le Y(t)\le\infty$ is mapped
bijectively into the interval $1\le W(t)\le\sqrt{3}$ where
$Y(t)=1$ is mapped to the logarithmic branch point at $W(t)=1$
while $Y(t)=\infty$ becomes $W(t)=\sqrt{3}$. Again the evolution
of the ghost cosmology starts off at some value $Y_i>1$ at initial
time $t_i=0$ and condenses towards $Y(t_c)\rightarrow 1$ at later
times. Since $Y(t_c)=1$ gives a logarithmic singularity in the
solution, we see that the condensation process again takes an
infinite amount of time, i.e.~we have to set $t_c=\infty$.
Moreover, we have to choose the solution with positive $\mt$ to
have a time-flow in positive direction. If the initial condition
is such that $Y_i\gg 1$ the solution becomes
\beqa
|\mt|t=2\ln\left|\bigg(\frac{\sqrt{3}-1}{W(t)-1}\bigg)
\bigg(\frac{W(t)+1}{\sqrt{3}+1}\bigg)\right|
-W(t)+\sqrt{3} \; .
\eeqa

\subsection{$d<1$}

Finally for the last case with $d<1$ one finds \beqa |\mt| t =
\frac{1}{\sqrt{1-d}}
\left(\arctan\Big(\frac{R(t)}{\sqrt{1-d}}\Big) -
\arctan\Big(\frac{R_i}{\sqrt{1-d}}\Big)\right) +
\frac{2}{\sqrt{d}} \ln\left|\frac{V(t)}{V_i}\right| \; , \eeqa
(sending $|\mt|t\rightarrow -|\mt|t$ also gives a valid solution,
however with opposite time arrow) where we have defined \beqa R(t)
= \frac{Y(t)+1-d}{\sqrt{3Y^2(t)-2Y(t)-1+d}} \; . \eeqa and initial
values $R_i,V_i$ at $t_i=0$. It is easy to check that $R(Y(t))$ is
a monotonically decreasing function along the interval $1\le
Y<\infty$ covering the range \beqa \frac{2-d}{\sqrt{d}}\ge
R(Y(t))>\frac{1}{\sqrt{3}} \; . \eeqa Therefore, the time $t$ is
monotonically decreasing with $Y$. Thus, with time running
forwards we have to evolve from large to small $Y$ values. At the
condensation point, $Y(t_c)=1$, we have $V(t_c)=\infty$ from which
we see that once again the condensation process of the ghost will
take an infinite amount of time, hence $t_c=\infty$.

\subsection{Transition from Radiation to Dark Energy Dominated Universe}

\begin{figure}[th]
\includegraphics[scale=0.7]{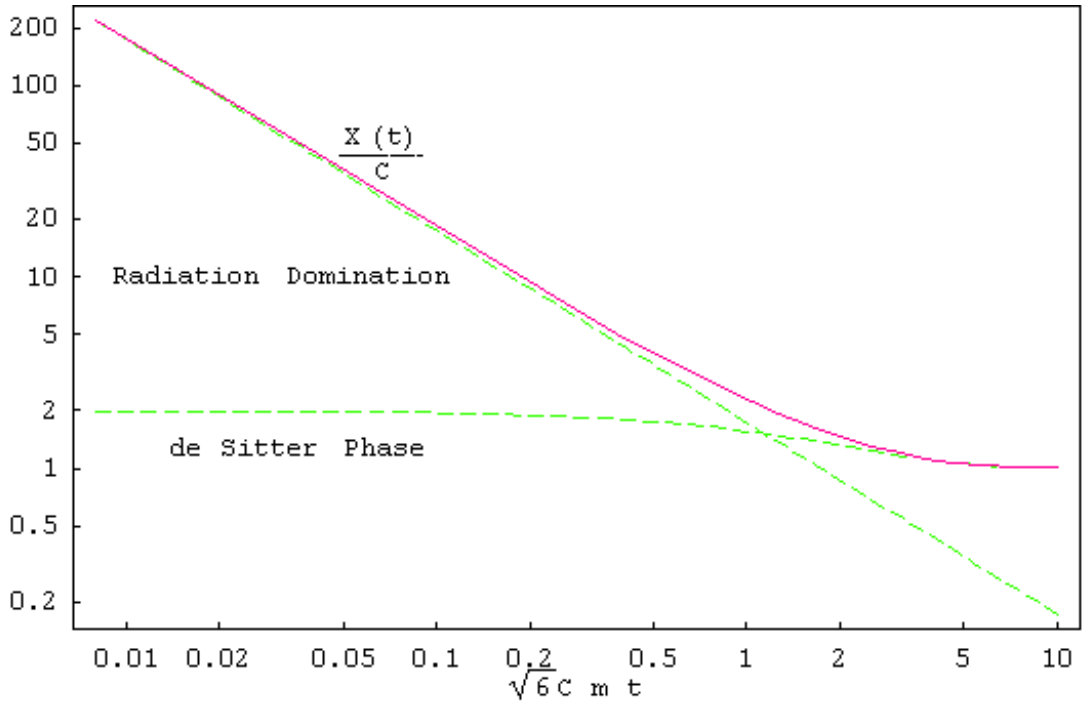}
\caption{\label{FigYPlot2} Plot of the ghost field $X(t)$ (upper
red curve) as a function of time for a ghost cosmology with $P(X)
= \frac{1}{2}(X-C)^2-D;\; D=\frac{2}{3}C^2$. One can clearly
see the transition from an early radiation dominated universe
(steep green dashed line) to a late time (dark) vacuum energy
dominated universe (flat green dashed line). The transition occurs
during $\sqrt{6}C m t\sim 0.1-5$.}
\end{figure}
To make these cosmological solutions more transparent, let us look
at their behavior at early and late times. By expanding the
solutions around $Y(t_i)=Y_i\gg 1$ at early times $t\rightarrow
t_i=0$ one finds for all the different cases that
\begin{alignat}{3}
X(t)&=\frac{M_{Pl}}{\sqrt{2}M^2t} \\
a(t)&\propto \left(\frac{M^2}{M_{Pl}}t\right)^{1/2} \; .
\end{alignat}
On the other hand at the late stages of the evolution upon
condensation of the ghost field where $Y(t)\rightarrow 1$ and
$t\rightarrow t_c=\infty$ one finds for all cases the general
behavior
\begin{alignat}{3}
X(t) &= C
\left(1+e^{-\sqrt{3D}\frac{M^2}{M_{Pl}}t}\right) \\
a(t) &\propto e^{\sqrt{\frac{D}{3}}\frac{M^2}{M_{Pl}}t} \; .
\end{alignat}
Here we have expressed these general results in the original
parameters. Hence all the obtained solutions describe irrespective
of the value of $D$ a {\em transition from an early radiation
dominated epoch to a dark energy dominated de Sitter cosmology at
late epoch}. At the same time this means a transition from
deceleration to acceleration as stated at the beginning of this
section.

\begin{figure}[th]
\includegraphics[scale=0.7]{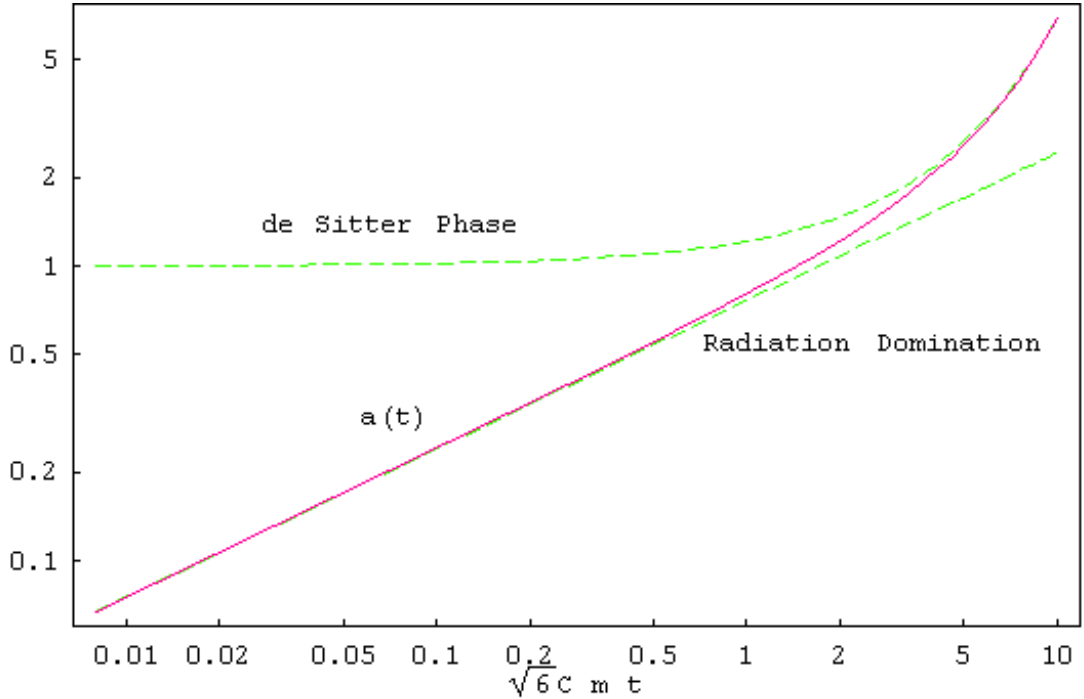}
\caption{\label{FigaPlot2} Plot of the scale-factor $a(t)$ (red
middle full curve) as a function of time for a ghost cosmology
based on $P(X) = \frac{1}{2}(X-C)^2-D;\; D=\frac{2}{3}C^2$. It
interpolates during the time interval $\sqrt{6}C m t\sim 0.2-7$
between a $t^{1/2}$ behavior (radiation dominated, lower green
dashed curve) at early times and a $\exp(\sqrt{D/3} mt)$ behavior
(vacuum energy dominated, upper green dashed curve) at late
times.}
\end{figure}
Again the behavior of these limiting cases can be quickly
understood by inspection of the density and pressure associated
with the ghost field and the related equation of state. From
(\ref{rhop}) we obtain that at early times
\beqa
\rho_\phi \rightarrow 3 p_\phi \; ,
\eeqa
while at late times
\beqa
\rho_\phi \rightarrow - p_\phi \; .
\eeqa
These are indeed the right equations of state showing that the
ghost field behaves like radiation at an early stage and like
vacuum energy at a late stage.

Let us now look in more detail at the complete transition
described by the given full solutions. Since all these solutions
show a very similar dependence, we will pick for concreteness the
easiest case of $d=4/3$ and choose its initial value $Y_i\gg 1$
which leads to a late time (dark) vacuum energy, $M^4 D =
\frac{2}{3}M^4C^2$. The explicit transition from the radiation
dominated to the vacuum energy dominated phase is depicted in
Fig.\ref{FigYPlot2} for $Y(t)$ and in Fig.\ref{FigaPlot2} for the
scale-factor $a(t)$. One observes that the transition occurs
during the time interval $\sqrt{6}Cmt=0.1-5$. We will see in
section VII that the current age of the universe estimated by $t_0
= 1/H_0$ leads to $\sqrt{6} Cm t_0 \sim 1$. The plots in
Fig.\ref{FigYPlot2} and Fig.\ref{FigaPlot2} show that this is {\em
right in the middle of the transition phase} - very similar to
what we have found before in section III in the dark matter case.

\section{From Large to Small Cosmological Constant?}

Both theories with quadratic function $P(X)$ which we have studied
in section III and V enjoyed a radiation dominated behavior with
characteristic $a(t)\sim t^{1/2}$ at early times. Moreover we have
seen in section IV that with a higher power $P(X)$ we can also
generate more general but exotic power-law expansions at an early
epoch. In view of the inflationary phase expected to have happened
in our universe at very early times, it will be also interesting
to ask whether such an inflationary phase could be described by a
suitable $P(X)$. Moreover, in view of the current astronomical
evidence for dark energy \cite{DarkEnergy}, one would like to join
it smoothly after various intermediate epochs again to a late time
accelerated de Sitter type expansion, however this time with a
drastically smaller vacuum energy. Irrespective of the
intermediate epochs which might differ from a de Sitter type
expansion, let us now -- by assuming that the ghost cosmology
framework applies way back to the times of inflation -- analyze
the structure of the corresponding $P(X)$. To describe a
transition between a de Sitter phase with large cosmological
constant and one with a small cosmological constant the
requirements on $P(X)$ are:
\begin{itemize}
\item $P(X)$ must possess (at least) two extrema, $X=C_1$ and
$X=C_2$ with $P'(C_1)=P'(C_2)=0$ at negative $P(C_1)<P(C_2)<0$.
The latter condition guarantees a positive vacuum energy at the
extremal points (to realize a very small late-time cosmological
constant, one would furthermore require a suitably small vacuum
energy $-P(C_2)M^4 \simeq \text{meV}^4$).
\end{itemize}
Given such a $P(X)$ we must moreover require that
\begin{itemize}
\item The cosmological evolution must run from $C_1$ -- the extremal
point with larger vacuum energy -- to $C_2$, the point with
smaller vacuum energy. This guarantees a transition from a large
positive cosmological constant $\Lambda_1=-P(C_1)M^4/M^2_{Pl}$ at
early times to a small cosmological constant $\Lambda_2 =
-P(C_2)M^4 / M^2_{Pl}$ at late times.
\end{itemize}
Two explicit classes of functions satisfying the first
requirements on $P(X)$ are the cubic (with positive parameters
$b,D>0$)
\beqa
P(X)=-(X-C)^3+b(X-C)^2-D
\eeqa
for which the critical points are
\begin{alignat}{3}
C_1 &= C \qquad\qquad &&\Rightarrow\qquad\qquad P(C_1) = -D \\
C_2 &= C+\frac{2}{3}b &&\Rightarrow\qquad\qquad P(C_2) =
-D+\frac{4}{27}b^3
\end{alignat}
and the Mexican hat kinetic function (with positive parameters
$g,D>0$)
\beqa
P(X)=-(X-C)^2+\frac{1}{2g^2}(X-C)^4-D
\eeqa
whose two relevant critical points (the left minimum and the
maximum in between the two minima) are
\begin{alignat}{3}
C_1 &= C-g\qquad\qquad &&\Rightarrow\qquad\qquad P(C_1) =
-D-\frac{1}{2}g^2 \\
C_2 &= C &&\Rightarrow\qquad\qquad P(C_2) = -D \; .
\end{alignat}
Both critical points are such that along $[C_1,C_2]$ the slope
$P'(X)$ is positive and also the remaining stability
(\ref{Constraint}) and positivity constraints (\ref{PosCon}) can
be satisfied along this interval. A more general $P(X)$ describing
various intermediate epochs might however be much more
complicated.

We will, however, argue that whatever the $P(X)$ satisfying the
above first requirement might be, no transition from large to
small cosmological constant is possible within the present
framework (based on a 3d flat metric Ansatz (\ref{FRW})). In order
to see this, let us start from the differential equation for $X$
(\ref{XODE}) and observe that $\ln|(P')^2 X|$ becomes $-\infty$ at
both $C_1$ as well as $C_2$. Hence, since the time evolution must
bring us from $C_1$ to $C_2$, the time-derivative of $\ln|(P')^2
X|$ has to start off at $C_1$ with a large positive value and end
at $C_2$ with a large negative value. Therefore, the lhs of
(\ref{XODE}) has to change sign in the interval $[C_1,C_2]$. Let
us now distinguish two cases. First, if $P'(X)\ge 0$ throughout
the interval $[C_1,C_2]$ then $\sqrt{2XP'-P}>0$ on the interval
and thus the rhs of (\ref{XODE}) cannot change sign as would be
required by the lhs. Thus no solution of (\ref{XODE}) exists for
the whole interval $[C_1,C_2]$. Second, in case that $P'(X)<0$ at
some point in the interval, (\ref{Constraint}) would be violated
and the cosmological evolution from $C_1$ to $C_2$ would suffer
instabilities against fluctuations along its way. We can therefore
conclude that in both cases no consistent solution exists which
would describe an evolution from $C_1$ to $C_2$ and thus a
transition from large to small cosmological constant. Notice that
this argument is valid for general $P(X)$ and thus cannot be
avoided by introducing a bunch of intermediate epochs -- as long
as the whole evolution from $C_1$ to $C_2$ is dominated by the
ghost.

\section{Evolution of Dark Energy/Matter in the Combined Ghost -
Radiation - Matter Cosmos}

Let us in this section study the cosmological evolution in the
presence of the ghost as well as additional ordinary matter and
radiation described by perfect fluids. With $n$ additional perfect
fluids the Einstein equation governing the behavior of the FRW
metric becomes
\beqa
G_{\mu\nu} = 8\pi G (T_{\phi,\mu\nu}+\sum_{i=1}^n T_{i,\mu\nu}) \;
,
\label{Einstein2}
\eeqa
with $T_{\phi,\mu\nu}$ the ghost energy-momentum tensor
(\ref{EMT1}) and
\beqa
T_{i,\mu\nu} = (\rho_i+p_i) u_\mu u_\nu + p_i g_{\mu\nu}
\eeqa
the energy-momentum tensor for the $i$th perfect fluid. A fluid
which is comoving in the FRW rest frame possesses a velocity
$u^\mu = (1,0,0,0)$ such that its energy-momentum tensor becomes
simply $T_{i,\mu\nu} = (\rho_i, p_i g_{rr}, p_i g_{\theta\theta},
p_i g_{\phi\phi})$.

The Einstein equation implies the local energy-conservation of the
combined ghost-fluid energy-momentum tensor. Motivated by the tiny
currently observed amount of 4\% ordinary baryonic matter as
opposed to 70\% dark matter \cite{DarkEnergy}, we will assume in
the sequel that the back-reaction of the perfect fluids on the
geometry is negligible as compared to the back-reaction of the
ghost which we will hold responsible in this section for all of
the current dark matter resp.~energy. That this assumption, once
true at the current epoch, remains valid at earlier epochs will be
shown explicitly below \footnote{It was pointed out by L. Sorbo that the
conventional Big Bang nucleosynthesis requires that non-standard
relativistic matter in the radiation era to be no more than 10\%. If
indeed
the conventional Big Bang nucleosynthesis scenario holds, this would be a
serious constraint on the ghost condensate proposal. However, the ghost to
ordinary matter coupling, which is outside the scope of our present paper,
could significantly alter the nucleosynthesis mechanism thereby
circumventing this constraint.}. It is thus reasonable
that the ghost and the fluids satisfy separate energy conservation laws
\beqa
\triangledown^\mu T_{\phi,\mu\nu} = 0 \; , \qquad
\triangledown^\mu T_{i,\mu\nu} = 0 \; .
\eeqa
As well-known, the fluid energy conservation law in an FRW
background is equivalent to
\beqa
\dot{\rho_i}+3H(\rho_i+p_i)=0 \; .
\eeqa
Together with an equation of state
\beqa
p_i = w_i\rho_i \; , \qquad w_i = const
\label{eos}
\eeqa
for the $i$th fluid one obtains for the mass density the solution
\beqa
\frac{\rho_i(t)}{\rho_{i,0}} =
\Big(\frac{a_0}{a(t)}\Big)^{3(1+w_i)} \; ,
\label{PowerLaw}
\eeqa
with $\rho_{i,0},a_0$ the present values. Similarly the ghost
energy conservation governs the ghost's dynamics as it is
equivalent to the ghost's equation of motion (\ref{Equ}) and is
solved by (\ref{GhostEOM2}). Of course the local
energy-conservation of the total mass density is just a
consequence of the Einstein equation and therefore not
independent.

The Einstein equation in the FRW background provides us with two
independent equations, the {\em Friedmann equation}
\beqa
H^2 = \frac{1}{3M_{Pl}^2}\Big( \rho_\phi + \sum_i\rho_i \Big)
\eeqa
and the evolution equation
\beqa
H^2+2\frac{\ddot{a}}{a} = -\frac{1}{M^2_{Pl}}\Big( p_\phi + \sum_i
p_i \Big) \; .
\eeqa
If we eliminate $H^2$ in this equation with help of the Friedmann
equation we arrive at the {\em acceleration equation}
\beqa
\frac{\ddot{a}}{a} = -\frac{1}{6M_{Pl}^2}\Big( \rho_\phi+3p_\phi
+ \sum_i(\rho_i+3p_i) \Big) \; .
\eeqa

The formal solution to the Friedmann equation is given by
\beqa
a(t)=a_0 e^{g(t,t_0)} \; , \qquad
g(t,t_0)=\pm\frac{1}{\sqrt{3}M_{Pl}} \int^t_{t_0}dt'\big(\rho_\phi
+ \sum_i\rho_i\big)^{1/2} \; .
\label{Scale2}
\eeqa
Plugging this into the acceleration equation to eliminate the
scale-factor leads to the equation
\beqa
\mp\frac{2M_{Pl}}{\sqrt{3}}
\frac{d}{dt}\big(\rho_\phi+\sum_i\rho_i\big)^{1/2}
= (\rho_\phi+p_\phi)+\sum_i(\rho_i+p_i) \; ,
\label{rhoEvol}
\eeqa
which determines the evolution of the densities once the equation
of states (\ref{eos}) are used together with (\ref{PowerLaw}) and
(\ref{Scale2}). The resulting differential equation is no longer
straightforward to solve. Fortunately, this is not needed since a
potential explanation of dark matter or dark energy at the current
epoch through the ghost would imply that today the ghost dominates
the fluid components by far and we will shortly see that this
feature also persists if we evolve the ghost-radiation-matter
system back to much earlier and also later times.

Nevertheless, before proceeding with the full combined
ghost-radiation-matter system let us look at two important
limiting cases. The first is the limit of {\em early times} for
the theory given by $P(X) = \frac{1}{2}(X-C)^2 - D$. At early
times we know that the baryonic matter contribution is negligible
as compared to radiation. The latter has an equation of state
$\rho_R = 3p_R$. Moreover at these early times we have found that
for this $P(X)$ the ghost also behaves like radiation with
$\rho_\phi=3p_\phi$. Therefore in this regime (\ref{rhoEvol}) can
be easily solved by (choosing the relevant minus sign)
\beqa
t\rightarrow 0\, : \qquad \rho_\phi+\rho_R =
\frac{3M_{Pl}^2}{4t^2}
\eeqa
leading via (\ref{Scale2}) to
\beqa
t\rightarrow 0\, : \qquad a(t)\propto t^{1/2} \; .
\eeqa

The second interesting limit is for {\em late times} for the
theory defined by the kinetic function $P(X)=\frac{1}{2}(X-C)^2$.
This kinetic function did not lead to dark energy at late times
but instead to a non-relativistic dark matter behavior with
$\rho_\phi\gg p_\phi$. Furthermore we know that at late times
ordinary matter dominates over radiation such that we have here
for the perfect fluid an equation of state $\rho_B \gg p_B$ as
well. Also in the late time limit (\ref{rhoEvol}) is easy to solve
through (again with the relevant minus-sign)
\beqa
t\rightarrow t_c \, : \qquad \rho_\phi+\rho_B =
\frac{4M_{Pl}^2}{3t^2}
\eeqa
which via (\ref{Scale2}) leads to the corresponding late time
scale-factor behavior
\beqa
t\rightarrow t_c \, : \qquad a(t)\propto t^{2/3} \; .
\eeqa

\subsection{Time-Dependence of Ghost Dark Matter, Radiation and Matter}
\begin{figure}[t]
\includegraphics[scale=0.7]{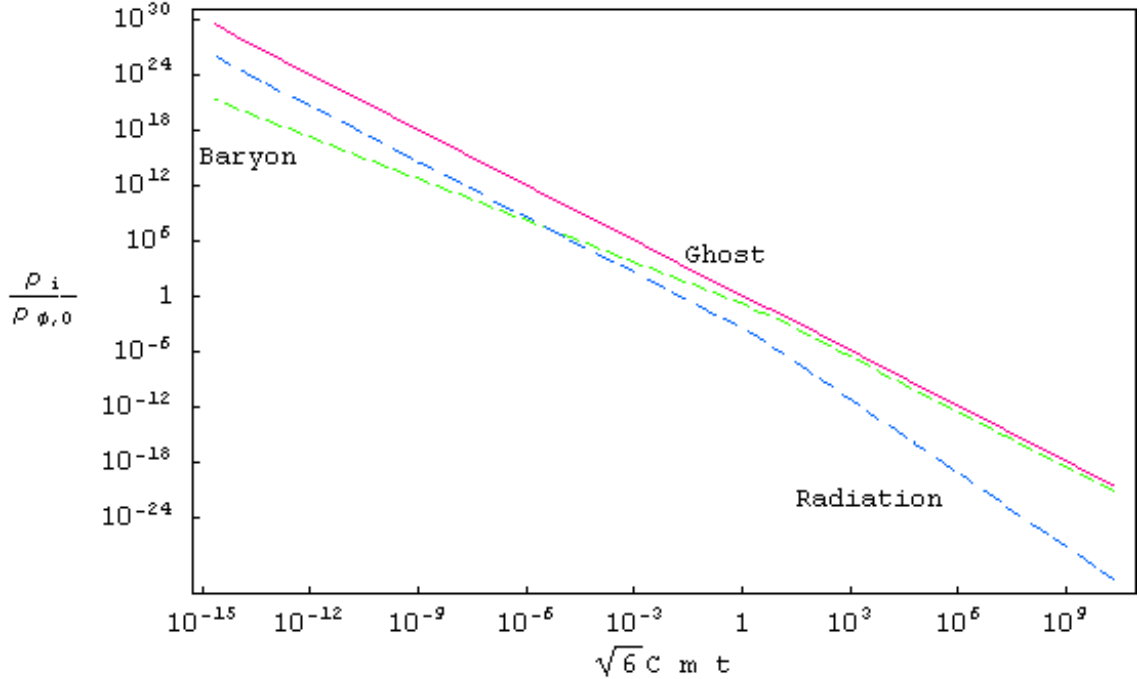}
\caption{\label{FigDMPlot} Time dependence of the mass densities
$\rho_\phi, \rho_B, \rho_R$ for the ghost scalar (which is assumed
to account for the present dark matter), baryonic matter and
radiation in a ghost cosmology based on $P(X)=\frac{1}{2}(X-C)^2$.
If the ghost accounts for the dark matter today then it dominates
ordinary matter and radiation not only today but even more so in
the past as well.}
\end{figure}

Let us now study the time evolution of ghost dark matter, baryonic
matter and radiation. We will therefore take the theory with
$P(X)=\frac{1}{2}(X-C)^2$ in which the ghost led to
non-relativistic matter at a late epoch. We know from observation,
that nowadays baryonic visible matter contributes 4\% and
radiation only $0.008$\% to the total mass density of the universe
as compared to 26\% coming from dark matter \cite{DarkEnergy}. We
will here equate the latter with the ghost dark matter. It will
therefore be a good approximation at the current epoch to neglect
the backreaction of ordinary baryonic matter and radiation against
that of the ghost. In this ghost dominated background we will now
evolve baryonic matter and radiation according to the simple
power-law behavior (\ref{PowerLaw}) with equation of state
parameters $w_B=0$ and $w_R=\frac{1}{3}$ and scale-factor $a(t)$
given by (\ref{a1}). We will shortly see that the dominance of the
ghost over baryonic matter and radiation becomes even more
pronounced when we look back in time. Also at late times the
dominance of the ghost will be preserved. This will then establish
the relevance of our cosmological solution (\ref{Y1}), (\ref{a1})
which took only the backreaction of the ghost into account.

We still have to fix the relative normalizations of the densities
which we will do at the current epoch. To this end, let us
estimate the age of the universe by the current Hubble parameter,
$t_0 = 1/H_0 \simeq 1/(10^{-33} \text{eV})$. To obtain a current
dark matter density of order $\text{meV}^4$ we have to adjust the
current ghost mass density $\rho_\phi=2C^2M^4(Y_0-1)$ at this
order (here $Y(t_0)=Y_0$). It is therefore natural (avoiding a
finetuning of $Y_0$) to set $CM^2\sim\text{meV}^2$. With this
information we arrive at a value for $\mt t_0 = \sqrt{6} C m t_0$
which is close to $1$. Consequently $Y_0=2.32$ and the
time-dependence of the scale-factor (\ref{a1}) becomes
\beqa \frac{a_0}{a(t)} = \frac{Y(t)^{1/6}(Y(t)-1)^{1/3}}{1.26} \; .
\label{aNorm} \eeqa
Combining this information with the present measured
densities \cite{DarkEnergy}, \cite{Verde} (the present radiative
contribution $\rho_{R,0}$ arises from the contribution of three
relativistic neutrinos plus the cosmic microwave background
photons \cite{Peebles} and the assumption of a Hubble parameter
$H_0=100\, h\, \text{km}\, \text{sec}^{-1}\, \text{Mpc}^{-1}$ with
$h=0.71$ \cite{Hubble})
\beqa
\rho_{\phi,0} \; : \; \rho_{B,0} \; : \; \rho_{R,0}
\quad
= \quad 26 \; : \; 4 \; : \; 8\times 10^{-3}
\eeqa
allows us to fix the relative normalization. Thus, the properly
normalized mass densities become
\begin{alignat}{3}
\frac{\rho_\phi}{\rho_{\phi,0}} &=  \frac{(Y(t)-1)(3Y(t)+1)}{10.27} \\
\frac{\rho_B}{\rho_{\phi,0}} &= \frac{4}{26} \times \Big(\frac{a_0}{a(t)}\Big)^3 \\
\frac{\rho_R}{\rho_{\phi,0}} &= \frac{8\times 10^{-3}}{26} \times
\Big(\frac{a_0}{a(t)}\Big)^4 \; .
\end{alignat}
with $\rho_{\phi,0}\equiv \rho_{DM,0}$ the current dark matter
density. Together with (\ref{aNorm}) this gives all densities as
functions of $Y(t)$ and thus via the inverted solution (\ref{Y1})
as functions of cosmic time $t$. The time-dependence of
$\rho_\phi,\rho_B,\rho_R$ is plotted in Fig.\ref{FigDMPlot}. We
see once more that the ghost behaves like radiation at early times
and assumes a non-relativistic matter-like behavior at late times.
The most important aspect of Fig.\ref{FigDMPlot}, however, is that
the {\em ghost's domination over all other contributions persists
to earlier epochs and becomes even more pronounced at earlier
times}. At later epochs the dominance of the ghost will be
preserved as well. Our approximation to choose (\ref{Y1}) as the
background at all times, not only at the present epoch, proves
therefore to be justified and establishes that our simple
cosmological solution (\ref{Y1}), (\ref{a1}) is actually the
relevant one also in the presence of ordinary matter or radiation.

To obtain the normalized mass densities $\Omega_\phi =
\rho_\phi/\rho_c$ and $\Omega_i = \rho_i/\rho_c$, obeying
$\Omega_\phi+\Omega_B+\Omega_R=1$, one would have to divide
further by the critical density $\rho_c=3M_{Pl}^2H^2$. Since we
use the approximation that the ghost alone determines the
geometry, i.e.~that $\rho_c = \rho_\phi+\rho_B+\rho_R \simeq
\rho_\phi$, it follows that $\rho_c$ resp.~$H$ (given by
(\ref{HDM})) are known functions of $Y(t)$ as well. Thus the
solution (\ref{Y1}) fully specifies the evolution of all
$\Omega_\phi,\Omega_i$. In particular in this approximation, we
will have $\Omega_\phi \simeq 1$ at any time.

\subsection{Time-Dependence of Ghost Dark Energy, Radiation and Matter}
\begin{figure}[t]
\includegraphics[scale=0.7]{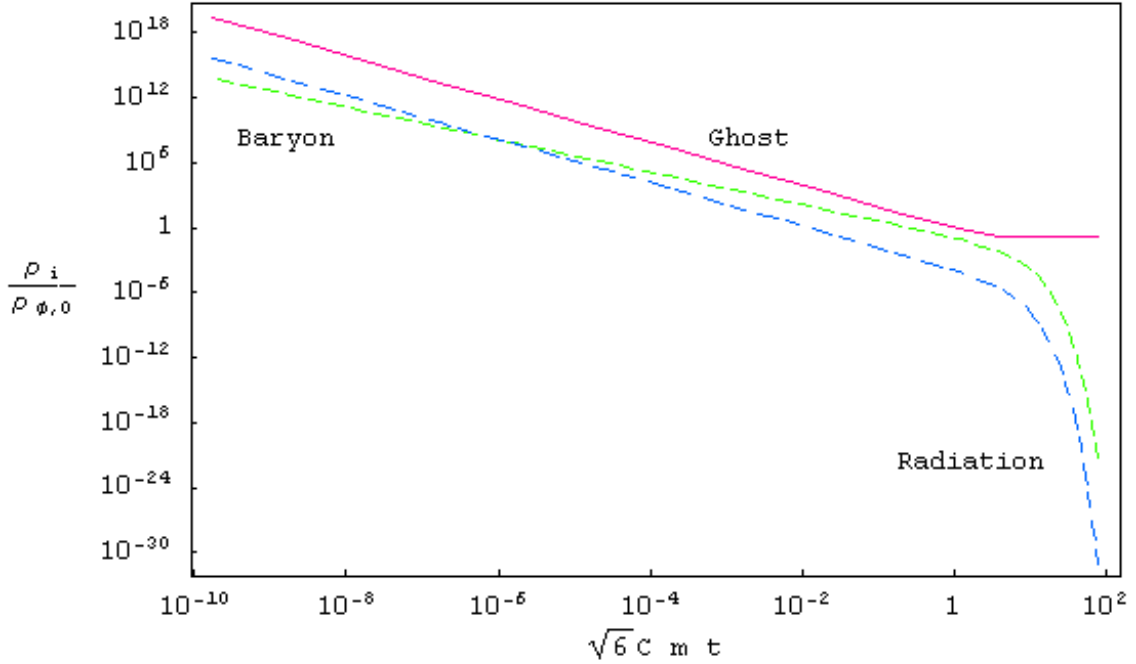}
\caption{\label{FigDEPlot} Time dependence of the mass densities
$\rho_\phi,\rho_B,\rho_R$ for the ghost field (which is identified
here with the dark energy), baryonic matter and radiation. Close
to the present age of the universe, where $\sqrt{6}Cmt_0\simeq 1$,
we find that the ghost field switches from a radiative phase to a
dark energy phase.}
\end{figure}

Let us next use the ghost field to account not for dark matter but
instead for the full amount of dark energy currently observed. To
this end, we have to take the theory with ghost kinetic function
$P(X)=\frac{1}{2}(X-C)^2-D$ leading to a late-time vacuum energy
$M^4 D$ when the ghost condenses. Again, from observation,
radiation and baryonic matter densities are negligible at the
current epoch as compared to the 70\% contribution to the mass
density coming from dark energy which we equate with the current
ghost mass density $\rho_{\phi,0}$. We are therefore entitled to
neglect all but the backreaction of the ghost on the geometry
which means that we have at present an FRW geometry with scale
factor given by one of the solutions in section V. Since all of
them behave similarly let us take the explicit solution (\ref{a4})
for $D=\frac{2}{3}C^2$ with initial condition $Y_i\gg 1$. If the
ghost field accounts for all of the present dark energy, which we
will assume in this subsection, we have to require that $M^4 D\sim
M^4C^2$ is of order $\text{meV}^4$.

Let us be more precise about the constants appearing in the
scale-factor and the mass densities. To this end we will again
estimate the current age of the universe by $t_0 = 1/H_0 \simeq
1/(10^{-33} \text{eV})$ which brings $\mt t_0 = \sqrt{6} C m t_0$
close to $1$. With knowledge of $t_0$ we can then fix the overall
constant in the scale-factor $a(t)$. Via (\ref{a4}) this leads to
\beqa \frac{a(t)}{a_0} = \frac{1}{0.80} \sqrt{1-e^{-\sqrt{2}Cmt}}
e^{\sqrt{2}Cm t/3} \; .
\eeqa
Moreover from (\ref{Y4}) we obtain that $Y_0=2.28$. Next, we fix
the relative normalizations of the mass densities through the
current observed mass densities for dark energy (which in this
subsection will be identified with the ghost mass density),
baryonic matter and radiation \cite{DarkEnergy}
\beqa
\rho_{\phi,0} \; : \; \rho_{B,0} \; : \; \rho_{R,0}
\quad
= \quad 70 \; : \; 4 \; : \; 8\times 10^{-3} \; .
\eeqa
This gives us the properly normalized mass densities ($d=4/3$)
\begin{alignat}{3}
\frac{\rho_\phi}{\rho_{\phi,0}} &=
\frac{\big((Y(t)-1)(3Y(t)+1)+d\big)}{11.37}
=  \frac{(3Y(t)-1)^2}{34.11} \\
\frac{\rho_B}{\rho_{\phi,0}} &= \frac{4}{70} \times
\Big(\frac{a_0}{a(t)}\Big)^3 \\
\frac{\rho_R}{\rho_{\phi,0}} &= \frac{8\times 10^{-3}}{70} \times
\Big(\frac{a_0}{a(t)}\Big)^4 \; .
\end{alignat}
as functions of $Y(t)$ where $\rho_{\phi,0}=\rho_{DE,0}$ is the
currently observed dark energy density. With the solution for
$Y(t)$ (\ref{Y4}) we finally obtain the full time dependence for
these densities which is plotted in Fig.\ref{FigDEPlot}.

We see from Fig.\ref{FigDEPlot} that the hierarchy among the three
contributions, which exists today if the ghost accounts for all of
dark energy, extends to earlier and later epochs likewise. We
therefore see that the assumption {\em to neglect the backreaction
of radiation and matter on the FRW geometry and to take only the
backreaction of the ghost into account is a valid assumption at
all epochs}. Herein lies the importance of the cosmological
solutions obtained in section V. Around $\sqrt{6}Cm t_\ast\sim
5\times 10^{-6}$ we reach the point where the baryonic matter and
radiation densities coincide. At $t=t_\ast$ the ghost field has
already adopted a radiative behavior as it runs with the same
slope as the radiation density. At even earlier times radiation
will dominate baryonic matter as in the standard hot big bang
cosmology but both are still dominated by far by the ghost's mass
density. Notice that while both $\rho_B$ and $\rho_R$ have the
same dependence on the scale-factor in the dark matter case (last
subsection) and the dark energy case (this subsection),
nevertheless their dependence on time is different in both cases
since the scale-factor $a(t)$ has a very different time-dependence
in these two cases. Let us mention that for the normalized
densities $\Omega_\phi, \Omega_B, \Omega_R$ the same remarks as in
the preceding subsection apply. In particular we have
$\Omega_\phi\simeq 1$ at all epochs.

The plot in Fig.\ref{FigDEPlot} reveals another interesting aspect
concerning the coincidence problem. The coincidence problem arises
in standard cosmology when the currently observed dark energy is
explained through an ordinary cosmological constant. The matter
density associated with the cosmological constant stays constant
during cosmic evolution. Thus, though the cosmological constant
would be the dominant energy contribution at present, this quickly
changes once one goes back to earlier times. Hence, the current
observation that the dark matter, the cosmological constant and to
a lesser extent also the matter contribution are of the same order
of magnitude present the coincidence problem. Notice that this
kind of {\em coincidence problem is much more harmless in the
ghost cosmology framework if we attribute dark energy completely
to the ghost}. Though at late times the ghost acts like a
cosmological constant its mass density grows towards earlier
times, keeping the hierarchy among the mass densities currently
observed more or less intact. Therefore, this hierarchy is not
specific to a finetuned snapshot of today's universe. The real
question which remains is to explain the hierarchy {\em at some
point in time} (see e.g.~\cite{KCC}). The time-dependence of the
ghost's mass density should have interesting cosmological
consequences for the verification or falsification of the ghost
gravity proposal.

\subsection{Describing both Dark Matter and Dark Energy by the
Ghost}
\begin{figure}[t]
\includegraphics[scale=0.7]{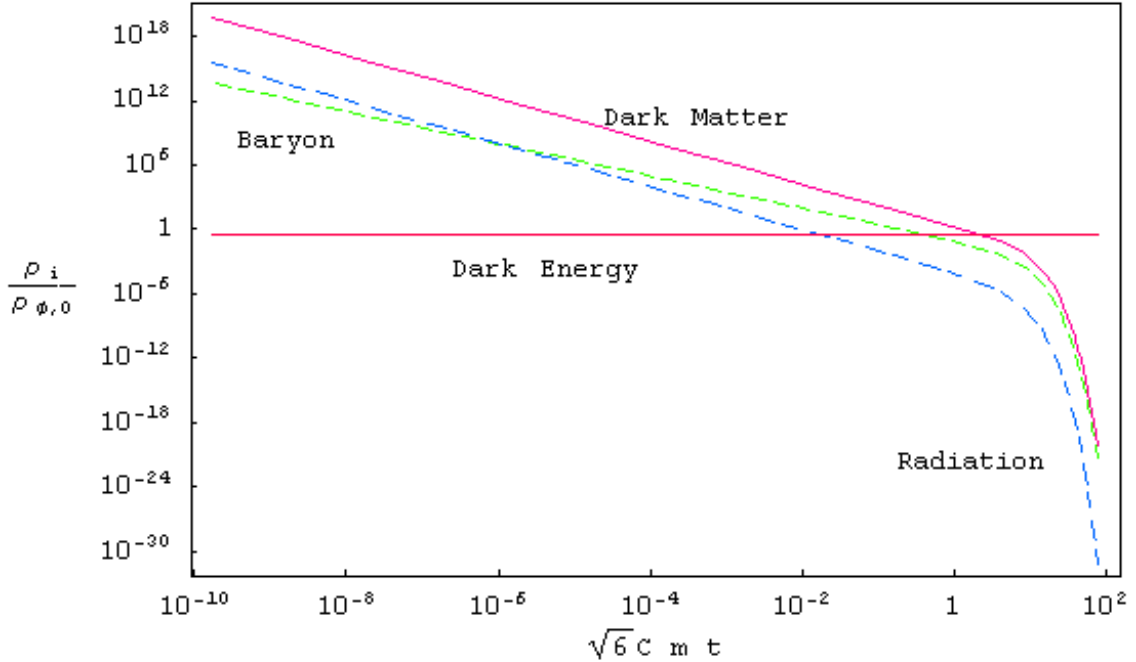}
\caption{\label{FigDEDMPlot} Time dependence of the mass densities
$\rho_{DE},\rho_{DM},\rho_B,\rho_R$ for the ghost field (which is
identified here with both the dark energy and dark matter),
baryonic matter and radiation. Close to the present age of the
universe, where $\sqrt{6}Cmt_0 = 1.674$, we find that the ghost
field switches from a radiative phase to a dark energy phase.}
\end{figure}

Up to now we have considered the dark matter and dark energy cases
separately by assuming that the ghost accounts for all of dark
matter or all of dark energy today. This was motivated by the
ghost's behavior towards late times. However, we have seen that
actually at the present epoch we are right at the transition
region from the early to the late time asymptotics. Therefore, the
ghost cosmology based on $P(X)=\frac{1}{2}(X-C)^2-D$ is indeed
capable of describing both dark matter and dark energy at the same
time as we will now discuss.

For this, let us split the ghost's mass density into a dark matter
part, $\rho_{DM}$, and a dark energy part, $\rho_{DE}$
\beqa \rho_\phi = \rho_{DM} + \rho_{DE} \eeqa where \beqa
\rho_{DM} = \frac{1}{2}M^4(X-C)(3X+C) \; , \qquad \rho_{DE} = M^4
D \; . \eeqa The associated pressures are \beqa p_{DM} =
\frac{1}{2}M^4(X-C)^2 \; , \qquad p_{DE} = -M^4 D \; . \eeqa Thus
the dark energy component is nothing but a standard cosmological
constant. Its evolution is therefore trivial and can be fixed once
and for all by identifying $\rho_{DE}$ with the current observed
dark energy density $\rho_{DE,0}$. This gives a value of
$Y(t_0)=1.614$ which translates into  $\mt t_0 = \sqrt{6} C m t_0
= 1.674$. Notice that we are using a different method of arriving
at the $Y(t_0)$ and $\mt t_0$ as we have an additional quantity to
fit. We find that this value of $\mt t_0$ coincides with our
previous estimates which have shown that this value is of order
one. On the other hand, the dark matter component is the same as
in subsection A, only that the normalization is slightly
different. The splitting that we have done to match the current
dark matter and dark energy, though an important step in obtaining
a realistic cosmology, leads us back to the coincidence
problem for the constant $\rho_{DE}$.

\section{Jeans-Instability}
\begin{figure}[t]
\includegraphics[scale=0.7]{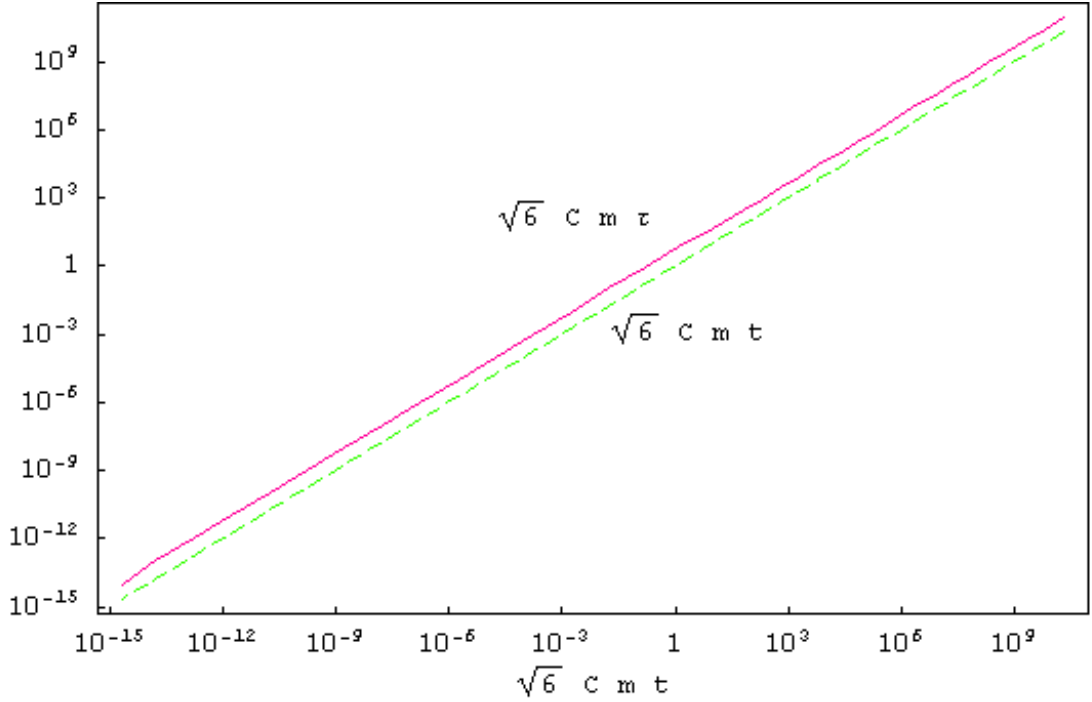}
\caption{\label{FigDMJnPlot} Plot of cosmic time $t$ and of the
free-fall time $\tau$ as functions of time for the ghost cosmology
based on $P(X)=\frac{1}{2}(X-C)^2$. At all times we have $\tau >
t$ indicating that no Jeans instability arises in the
corresponding cosmological solutions of section III.}
\end{figure}

\begin{figure}[t]
\includegraphics[scale=0.7]{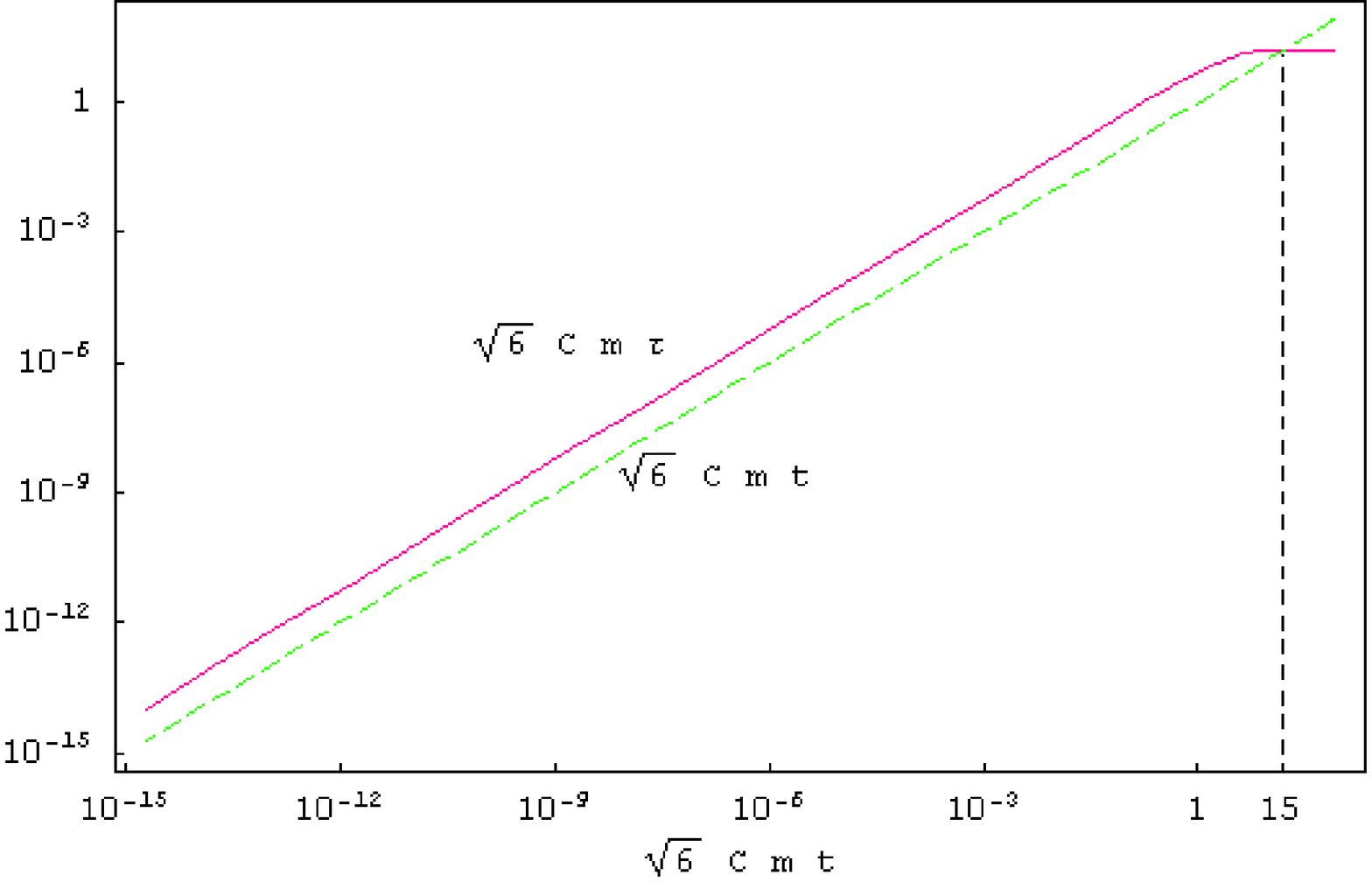}
\caption{\label{FigDEJnPlot} Plot of cosmic time $t$ and of the
free-fall time $\tau$ as functions of time for the ghost cosmology
based on $P(X)=\frac{1}{2}(X-C)^2-D$, ($D=\frac{2}{3}C^2$ is used
for the plot). At early times we find $\tau
> t$ indicating that no Jeans instability arises. However at late
times we see that there is a crossover to a regime with $\tau< t$.
Hence a Jeans instability occurs at late times for the
cosmological solutions of section V.}
\end{figure}

Let us finally analyze the stability of the obtained ghost
cosmological solutions in section III and V under gravitational
collapse, i.e.~Jeans instability. A universe with mass density
$\rho$ is unstable under perturbations if the perturbations have a
wavelength $\lambda$ greater than the critical Jeans wave length
\beqa \lambda_J = \frac{v_s}{\sqrt{G\rho}} \; . \eeqa We neglect
here factors of order one on the rhs which depend on the detailed
geometrical properties of the perturbed equilibrium state and the
perturbations. The isothermal speed of sound is denoted by $v_s$.
The corresponding time scale measuring the total time of the
ensuing collapse if $\lambda>\lambda_J$ is given by the free-fall
time \beqa \tau = \frac{1}{\sqrt{G\rho}} =
\frac{\sqrt{8\pi}M_{Pl}} {\sqrt{\rho}}\; . \eeqa
Thus, as long as
the cosmic time $t$ satisfies $t<\tau$, there was not sufficient
time for the universe to have collapsed and it will be stable. To
the contrary, for $t>\tau$ the universe had enough time to
collapse and will suffer from a Jeans instability. For the ghost
cosmologies based on $P(X)=\frac{1}{2}(X-C)^2-D$ we have to set
$\rho=\rho_\phi$ with \beqa \rho_\phi =
M^4\Big(\frac{1}{2}(X-C)(3X+C)+D\Big) =
\frac{C^2M^4}{2}\Big((Y-1)(3Y+1)+d\Big) \; . \eeqa

Let us first examine the dark matter case, i.e.~the theory with
$D=0$. The comparison between $t$ (from \ref{Y1}) and the
free-fall time $\tau$ is presented in Fig.\ref{FigDMJnPlot}. We
see that in this case no Jeans instability arises at any time. To
understand this better let us investigate the early and late time
limits. For the ghost mass density these are \beqa \rho_\phi
\rightarrow \left\{
\begin{array}{lll}
t\rightarrow 0 & : & \;\; \frac{3}{2}M^4X^2 \\
t\rightarrow \infty & : & \;\; 2CM^4 (X-C)
\end{array}
\right.
\eeqa
This leads to the following expressions for the free-fall time
\beqa
\tau \rightarrow
\left\{
\begin{array}{lll}
t\rightarrow 0 & : & \;\; \frac{4\sqrt{\pi}}{\sqrt{3}mX} \\
t\rightarrow \infty & : & \;\; \frac{2\sqrt{\pi}}{m(X-C)}
\end{array}
\right.
\eeqa
On the other hand, the cosmic time in these limits can be found
from the expressions in section III as
\beqa
t \rightarrow
\left\{
\begin{array}{lll}
t\rightarrow 0 & : & \;\; \frac{1}{\sqrt{2}mX} \\
t\rightarrow \infty & : & \;\;
\frac{\sqrt{2}}{\sqrt{3C}m\sqrt{X-C}}
\end{array}
\right.
\eeqa
Obviously in both limits, we find that $\tau > t$. Notice that in
the late time limit the ghost condenses, $X\rightarrow C$.

For the dark energy case with nonvanishing $D\ne 0$ we see,
however, from the plot in Fig.\ref{FigDEJnPlot} that a Jeans
instability arises at late times (the actual time when this
happens depends on $D$ but is always in the future, i.e.~later
than $t_0$). Let us again look at the limiting expressions for
this case. We find for the mass densities \beqa \rho_\phi
\rightarrow \left\{
\begin{array}{lll}
t\rightarrow 0 & : & \;\; \frac{3}{2}M^4X^2 \\
t\rightarrow \infty & : & \;\; M^4 D
\end{array}
\right.
\eeqa
which leads to the following free-fall times at early and late
times
\beqa
\tau \rightarrow
\left\{
\begin{array}{lll}
t\rightarrow 0 & : & \;\; \frac{4\sqrt{\pi}}{\sqrt{3}mX} \\
t\rightarrow \infty & : & \;\; \frac{\sqrt{8\pi}}{m\sqrt{D}}
\end{array}
\right.
\eeqa
To compare, we find for the cosmic time from the expressions in
section V that
\beqa
t \rightarrow
\left\{
\begin{array}{lll}
t\rightarrow 0 & : & \;\; \frac{1}{\sqrt{2}mX} \\
t\rightarrow \infty & : & \;\; -\frac{\ln(X-C)}{\sqrt{3D}m}
\end{array}
\right.
\eeqa
Therefore, at early times we find $\tau > t$ and consequently no
Jeans-instability can occur. However, at late times we clearly have
 $t>\tau$ due to the logarithmic divergence in $t$
indicating a Jeans-instability at a late epoch. Notice that the
precise time, $t_{crossover}$, when the Jeans instability will set
in depends on $D$. E.g.~in the case of $D=\frac{2}{3}C^2$ one
finds that $\sqrt{6}Cm t_{crossover}=15.04.$ Since for
$D\rightarrow 0$ we must recover the previously analyzed case
which had no Jeans-instability at any time, we recognize that
sending $D\rightarrow 0$ implies sending the crossover time, at
which $\tau = t_{crossover}$, towards infinity. Thus for small
enough $D$, the Jeans instability can be pushed far into the
future. Notice that the Hubble friction which is present when the
instability sets in, will cure it. 

Additionally, higher derivative terms will not affect the stability. At
early times, the higher derivative fluctuations are suppressed with
respect to $P(X)$ while at late times, $P(X)$ is vanishing and hence
higher derivative fluctuations might play a part. However, the detailed
fluctuation analysis done in \cite{ACLM} shows that we do have a stable
background.

\section{Conclusion}

We have derived the exact cosmological solutions for a universe
composed primarily of a ghost condensate. It was further shown
that these solutions would smoothly interpolate between different
standard cosmologies. We find, generically for quadratic kinetic
functions, an early-time radiation-dominated universe which
transitions into a de-Sitter expansion phase or a late-time
matter-dominated scenario depending on whether the cosmological
constant is present or not. This could also be understood in terms
of the stress energy tensor of the ghost which gives rise to an
equation of state that clearly exhibits early-time radiation-like
and late-time dust-like behavior or de Sitter acceleration. For
higher order kinetic functions we find more exotic power law FRW
universes. An important result is that, given the current age of
the universe in terms of the Hubble time, we are placed right at
the transition regime. Moreover, the cosmological solutions show
that the ghost condensation process takes an infinite amount of
time. This preserves us from entering the regime of unstable vacua
in finite cosmic time.

As a first step towards constructing fully realistic cosmologies,
we considered the evolution of the universe by extrapolating from
present dark energy/matter, baryonic and radiation densities to
earlier times. We find that the ghost contribution completely
dominates at all times and hence must be the principal driving
force of cosmological evolution. Once we adjust for the dark
matter/energy densities observed today by assuming that the ghost
can account for these, our solutions map out the entire evolution
of these densities. The other important consideration is the Jeans
instability and we have shown that for the ghost dark matter case
no such instability arises. In contrast, the dark energy case
exhibits a late-time Jeans instability. It occurs when the
universe has already entered the de Sitter expansion phase which
means that Hubble friction should cure the instability.

\begin{acknowledgments} T
he authors are grateful to Markus A. Luty and
Lorenzo Sorbo for interesting discussions. We have been supported by the
National Science Foundation under Grant Number PHY-0354401. During the
final stages of this project, S.P.N. was supported by the Department Of
Energy under contract DE-FG02-91ER40626. 
\end{acknowledgments}

\end{document}